\documentclass[conference]{IEEEtran}
\IEEEoverridecommandlockouts
\IEEEpubid{
\begin{minipage}{\textwidth}\centering
\vspace{7em}
\footnotesize
\copyright\ \the\year{} IEEE. Personal use of this material is permitted. Permission from IEEE must be obtained for all other uses, in any current or future media, including reprinting/republishing this material for advertising or promotional purposes, creating new collective works, for resale or redistribution to servers or lists, or reuse of any copyrighted component of this work in other works.
\end{minipage}
}

\usepackage{cite}
\usepackage{amsmath,amssymb,amsfonts}
\usepackage{algorithmic}
\usepackage{graphicx}
\usepackage{textcomp}
\usepackage{subfigure}
\usepackage[table,xcdraw]{xcolor}
\usepackage{adjustbox}
\usepackage{multirow}
\usepackage{makecell}
\usepackage[hidelinks]{hyperref}
\usepackage{enumitem}
\usepackage{amsmath}
\usepackage{booktabs} 
\usepackage{multicol}
\usepackage{multirow}
\def\BibTeX{{\rm B\kern-.05em{\sc i\kern-.025em b}\kern-.08em
    T\kern-.1667em\lower.7ex\hbox{E}\kern-.125emX}}

\begin{document}
   




    



\title{Securing WiFi Fingerprint-based Indoor Localization Systems from Malicious Access Points\\}

\author{\IEEEauthorblockN{Fariha Tanjim Shifat, Sayma Sarwar Ela, Mosarrat Jahan}
\IEEEauthorblockA{Department of Computer Science and Engineering, University of Dhaka, Dhaka, Bangladesh \\ 
Email:2018-025-345@student.cse.du.ac.bd,2018-325-351@student.cse.du.ac.bd, mosarratjahan@cse.du.ac.bd}
}

\maketitle

\IEEEpubidadjcol

\begin{abstract}
WiFi fingerprint-based indoor localization schemes deliver highly accurate location data by matching the received signal strength indicator (RSSI) with an offline database using machine learning (ML) or deep learning (DL) models. However, over time, RSSI values degrade due to the malicious behavior of access points (APs), causing low positional accuracy due to RSSI value mismatch with the offline database. Existing literature lacks the detection of malicious  APs in the online phase and mitigating their effects. This research addresses these limitations and proposes a long-term, reliable indoor localization scheme by incorporating malicious AP detection and their effect mitigation techniques. The proposed scheme uses a Light Gradient-Boosting Machine (LGBM) classifier to estimate locations and integrates simple yet efficient techniques to detect malicious APs based on online query data. Subsequently, a mitigation technique is incorporated that updates the offline database and online queries by imputing stable values for malicious APs using LGBM Regressors. Additionally, we introduce a noise addition mechanism in the offline database to capture the dynamic environmental effects. Extensive experimental evaluation shows that the proposed scheme attains a detection accuracy above  95\% for each attack type. The mitigation strategy effectively restores the system’s performance nearly to its original state when no malicious AP is present. The noise addition module reduces localization errors by nearly 16\%. Furthermore, the proposed solution is lightweight, reducing the execution time by approximately 94\% compared to the existing methods.
\end{abstract}

\noindent\textbf{Cite as:}

\noindent F. T. Shifat, S. S. Ela and M. Jahan, ``Securing WiFi Fingerprint-based Indoor Localization Systems from Malicious Access Points," in \textit{2025 IEEE 7th Symposium on Computers \& Informatics (ISCI)}, Kuala Lumpur, Malaysia, 2025, pp. 380-387, doi: 10.1109/ISCI65687.2025.11167252.

\vspace{0.5em}
\noindent\textbf{Link to the final, published paper in IEEE Xplore:}\newline
     https://doi.org/10.1109/ISCI65687.2025.11167252

\section{Introduction}
With the rapid advancement of wireless communication technologies and the Internet of Things (IoT), location-based services have become indispensable in many application domains. Various localization mechanisms that operate either indoors or outdoors are employed to determine location. In outdoor environments, positions can be accurately determined using the Global Navigation Satellite System (GNSS). However, indoor positioning is challenging due to the attenuation of the Global Positioning System (GPS) signal caused by building structures and their materials \cite{recentadvances}. Accurate indoor location tracking is critical for several purposes, such as firefighter tracking under hazardous conditions and in emergencies, location-based smart resource management, robotics, shopping mall navigation, healthcare, and virtual reality \cite{recentadvances}, \cite{3}, \cite{27}. Consequently, indoor localization has become a major focus for both academia and industry.

One of the most widely used and preferred techniques for indoor localization is WiFi-based fingerprinting. This method uses existing WiFi networks and does not require additional hardware installation \cite{31}. Existing research focuses mainly on designing a highly accurate WiFi fingerprint-based indoor localization system that minimizes localization errors \cite{10, 11, 8}. However, the security and privacy issues of indoor localization systems have not been thoroughly explored. The absence of adequate security mechanisms can result in security attacks, such as jamming, spoofing, and sniffing \cite{11, security_tiku}. These attacks ultimately lead to inaccurate location estimation, raising significant concerns about the reliability of localization schemes. 

A significant security issue is the presence of malicious access points that generate false RSSI signals \cite{security_tiku}. Over time, APs may behave maliciously due to component malfunctions \cite{29, 30} and security attacks such as signal jamming \cite{security_tiku}, spoofing \cite{11}, distance fraud \cite{dist_fraud-1}, \cite{dist_fraud-2}, and physical attacks \cite{10}. These malicious APs mislead localization systems and reduce location accuracy. Unreliable location accuracy can have serious consequences in critical situations. For example, if a person needs immediate medical attention in a hospital, inaccurate location prediction of healthcare professionals and facilities can elevate health risks. Similarly, robots navigating in complex indoor environments with unreliable localization methods may cause damage to assets and potentially lead to injuries. Existing literature deals with malicious APs during the initial database construction phase (offline phase) by selecting a subset of feasible APs \cite{10}, \cite{11}. However, filtering out malicious APs only in the offline phase does not guarantee the long-term robustness of the indoor localization systems, as a reliable AP may become malicious over time. Current literature lacks research on detecting the long-term malicious behavior of APs \cite{30, 38, elm, 35}. 

After detecting malicious APs, it is crucial to mitigate their effect to ensure the accurate functioning of the indoor localization system. One approach to achieving this goal is to reconstruct the database. However, each time a malicious AP is detected, conducting a site survey to build the database in a dynamic environment is highly expensive and labor-intensive \cite{38}. Existing literature has explored the use of crowdsourcing techniques \cite{35} for creating and updating the fingerprint database, which significantly reduces human labor. However, this scheme is vulnerable to various security issues \cite{crowd_sec}. Additionally, robots are utilized for autonomous database construction, which can be expensive and may not perform effectively in challenging environments \cite{robot}. Instead, it would be more beneficial to develop a mechanism that enables an indoor localization system to function accurately despite the presence of a reasonable percentage of malicious APs, eliminating the need for frequent database reconstruction. This would enhance the system's accuracy while reducing the costs associated with repeated database creation. Although several existing research works mitigate the effect of RSSI value alteration due to environmental factors \cite{37}, \cite{35}, \cite{38}, none of them consider the malicious behavior of APs causing abrupt and dynamic changes to RSSI values.

This paper addresses the above-mentioned shortcomings and proposes lightweight mechanisms for detecting malicious APs over the long term during their operation. Additionally, it introduces a mitigation strategy to manage the impact of malicious APs while maintaining the accuracy of location prediction over an extended period, without requiring frequent database reconstruction.

\section{Related works} \label{related_work}
WiFi RSSI-based indoor localization schemes provide highly accurate location data at a low additional cost. However, the location predicted by these models often becomes erroneous due to the alteration of RSSI values caused by malicious AP behaviour \cite{10}, \cite {dist_fraud-1}, \cite{security_tiku}, changes in layout \cite{35}, removal of AP \cite{elm}, position change of AP \cite{elm}, \cite {35}, and environmental factors \cite{31,37, elm}. Existing literary works \cite{10,11,12,14} detect and filter malicious APs during the model training phase (offline phase) of indoor localization systems. Here, a subset of optimal AP is selected, and using these selected nonmalicious APs, a fingerprint database is constructed to train the machine learning model. The selection of correlated APs increases localization accuracy and reduces computational complexity. Wang et al. \cite{10} utilized interclass dispersion to calculate AP confidence and selected APs with high confidence. Ye et al. \cite{11} used the Pearson Correlation Coefficient (PCC), where a lower correlation value indicates greater susceptibility of an AP to being malicious. 
Chen et al. \cite{12} considered the resolution capability to select a set of APs, where APs are chosen based on information gain. Panja et al. \cite{14} used the Binary Particle Swarm Optimization (BPSO) to select a set of suitable APs. In contrast, our proposed scheme uses Spearman's Correlation Coefficient (SRCC) to select correlated APs in the offline phase. We use normalization, correlated AP selection, and noise addition as data preprocessing, where normalization speeds up the computational time, and noise addition makes the system robust to long-term signal variation due to dynamic environmental change. 

Detecting malicious APs only in the offline phase is not sufficient for long-term model reliability. Several works explore long-term model robustness challenges in the online phase. For example, Yan et al. \cite{37} proposed a crowdsourcing-based mechanism to handle RSSI signal value alteration in the online phase. This scheme utilizes a denoising autoencoder to address RSSI alteration caused by environmental dynamics. It requires extensive, high-quality crowdsourced data for training, and the model is updated frequently, which can be resource-intensive. Li et al. \cite{30} addressed signal variation caused by dynamic environmental change and temporal effects using disagreement-based semi-supervised learning. This scheme struggles to handle sudden and significant alterations in the environment, which can impact localization accuracy. Tiku and Pasricha \cite{31} utilized Siamese neural encoders to address RSSI signal variation without the need for model retraining. This scheme handles signal alteration caused by environmental changes, human movement, and the removal or replacement of AP. Significant changes in the environment can affect localization accuracy. Huang et al. \cite{38} addressed location accuracy degradation caused by environmental changes. This scheme utilizes a Gaussian Process Regression to create an offline database and incorporates the Marginalized Particle Extended Gaussian process that uses crowdsourced data to recursively update the database. However, this scheme is susceptible to localization errors due to rapid fluctuations in signal strength, and its effectiveness depends on the quality of the crowdsourced data. Jiang et al. \cite{elm} addressed RSSI signal alteration caused by the change in the number of AP and environmental factors. The authors proposed a feature-adaptive extreme learning machine-based approach that can adapt to a changing number of features. This scheme faces difficulties in coping with sudden environmental changes and runs the risk of overfitting to recent signal variations. Yang et al. \cite{35} proposed the Altered AP Identification and Fingerprint Updating (AAIFU) scheme, which detects altered APs in the online phase by considering changes in RSSI values due to positional changes of APs. This scheme uses a Gradient Boosting Decision Tree (GBDT) regression model to identify the positional changes of APs. It then updates the database by replacing the outdated RSSI values with fresh RSSI values obtained through GBDT regression.  However, the effectiveness of this scheme depends on a large volume of crowdsourced data. Moreover, positional changes of an AP only create persistent changes in RSSI values. Among these works, Li et al. \cite{30}, Huang et al. \cite{38}, Jiang et al. \cite{elm}, and Yang et al. \cite{35} update the database to keep the fingerprints fresh. On the other hand, none of the schemes, except Yang et al. \cite{35}, identify the causes of RSSI signal alteration. In contrast to the existing works, our proposed scheme addresses the RSSI value alteration caused by malicious AP behavior, characterized by abrupt and random changes in RSSI values. Our proposed scheme detects malicious APs in real time by analyzing data from various location queries. Additionally, it utilizes the LGBM regression model \cite{base-lgbm} to update the existing database, which mitigates the effect of malicious APs on the performance of the indoor localization scheme. Here to note that the proposed scheme handles both static and dynamic changes in RSSI values.

\section{System Model}\label{system_model}
\begin{figure}
	\centering
 \includegraphics[height=2 in, width=0.40\textwidth]{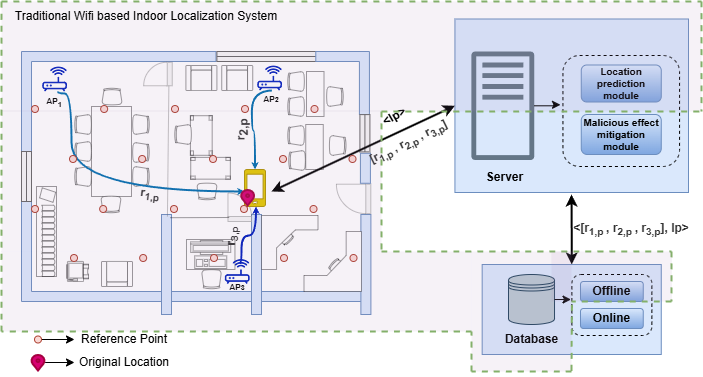}
 \caption{System model.}
\label{fig: system_model}
\end{figure}

As shown in Fig. \ref{fig: system_model}, the proposed scheme augments the existing WiFi fingerprint-based indoor localization system \cite{model} by including a \textit{malicious effect mitigation module} and \textit{online database}. The proposed system model comprises $q$ access points and $r$ reference points (RPs). The \textit{offline database} comprises a set of records of form $[RSSI_j, l_j]$, where $RSSI_j = [r_{i,j}]$ is the RSSI vector and $r_{i,j}$ is the RSSI value of $i$th AP measured at $j$th RP, and $l_j$ is the location of the $j$th RP, $i \in 1, 2, \ldots, q$, and $j \in 1, 2, \ldots, r$. In the \textit{online} phase, a device measures a fingerprint vector $[RSSI]$ with respect to its location $p$ and sends it to the \textit{location prediction module}. The \textit{location prediction module} matches $[RSSI]$ with the records stored in the \textit{offline database} using either an ML or DL model and returns the estimated location $l_p$ to the mobile device. The \textit{online database} is constructed by saving the results of online queries along with associated RSSI vectors and used to detect malicious APs in the \textit{online} phase. The \textit{malicious effect mitigation module} detects malicious APs and takes measures to diminish their effect. The proposed system model subjects to false fingerprints both in \textit{online} and \textit{offline} phases due to various environmental factors \cite{31, elm}, temporal effects \cite{11}, and malicious behavior of APs \cite{10, 11}. We assume that the \textit{databases}, the \textit{location prediction module}, and the \textit{malicious effect mitigation module} reside in a trusted server. 

\section{Proposed Scheme}\label{proposed_scheme}
\begin{figure}
    \centering
    \includegraphics[height=2 in, width=.45\textwidth]{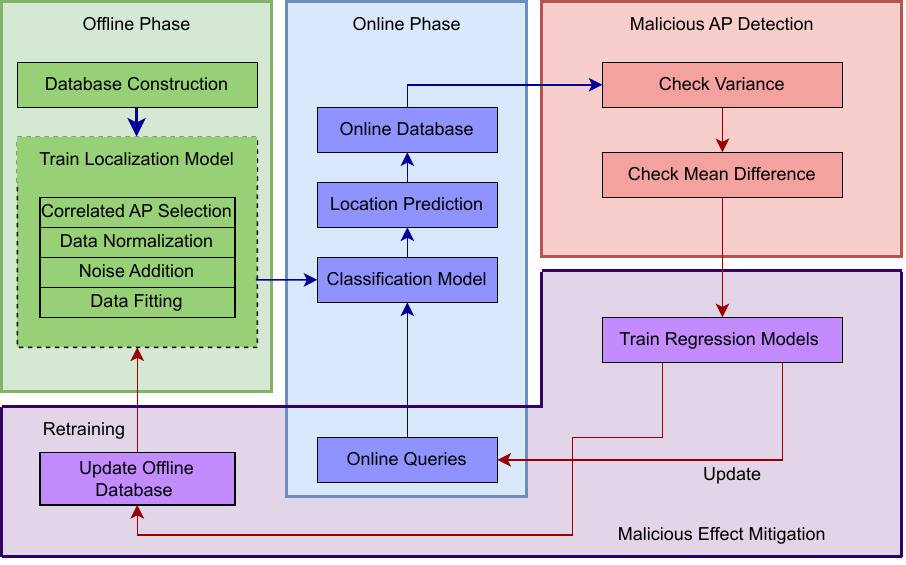}
    \caption{Overview of the proposed scheme.}
    \label{fig:proposed_scheme}
\end{figure}

As shown in Fig. \ref{fig:proposed_scheme}, the proposed scheme consists of four modules: \textit{offline phase}, \textit{online phase}, \textit{malicious AP detection}, and \textit{malicious effect mitigation}. The notations used to describe the proposed scheme are listed in Table \ref{tab:list_notations}.
\begin{table} 
\centering
\caption{List of Notations}
\scalebox{.95}{
\begin{tabular}{|l|p{6.5cm}|}
\hline
\textbf{Notation}                 & \textbf{Description}     \\ \hline
$r_{i,j}$  & RSSI value of $i$th AP at $j$th RP\\ \hline
$l_j$ & Location of the $j$th RP\\ \hline
$RSSI_j$   & A vector of form $[r_{i,j}]$ for $j$th RP and $i \in 1,2, \ldots,q$  \\ \hline
$RSSI'_j$   & An offline data sample of form $[r'_{i,j}]$ for $j$th RP and $i \in 1,2, \ldots,n$ and $n\leq q$  \\ \hline
$[RSSI'_j, l_j]$  & A labeled sample representing location $l_j$ for $RSSI'_j$ in offline database    \\ \hline   
$RSSI''_p$ & Online query of form $[r''_{i,p}]$ for $p$th RP where $p$ is unknown\\ \hline
$[RSSI''_p, l_p]$  & A labled sample representing location $l_p$ for $RSSI''_p$ in online database    \\ \hline   
\end{tabular}}
\label{tab:list_notations}
\end{table}

\subsection{Offline Phase}
\subsubsection{Database Construction} At first, a reliable database is created using the site survey \cite{radar}. This mechanism involves a person with multiple reliable mobile devices visiting every RP. It collects several labeled samples in the form $[RSSI_j, l_j]$, where $RSSI_j$ is an RSSI fingerprint vector computed at $j$th RP and $l_j$ is the location of the $j$th RP. The raw database created by the site survey is a collection of records $[RSSI_j, l_j]$.

\subsubsection{Correlated AP Selection}\label{correlated_AP}
The proposed scheme selects a set of reliable APs from the available ones in the environment using Spearman's Rank Correlation Coefficient (SRCC) \cite{Spearman} method with a threshold value set to 0.1 empirically, which is optimal for the considered UJI's dataset \cite{long-term}. Suppose the number of selected correlated APs is $n$. The proposed scheme constructs an offline database $\{[RSSI'_j, l_j]\}$ from the raw database constructed via site survey, where $RSSI'_j$ contains RSSI entries only for the selected $n$ APs.

\subsubsection{Data Normalization} \label{data_normal} The input RSSI values were scaled into [0,1] range using Eq. \ref{eqn:normalization} \cite{normalization}.
\begin{equation}
\small
r'_{i, j} = \begin{cases}
    0, & \text{if } r'_{i, j} = no\_signal \\
    \left(\frac{r'_{i, j} - \text{min}}{\text{max} - \text{min}}\right) \times (1 - 0.25) + 0.25, & \text{if } r'_{i, j}\neq no\_signal
    \end{cases}
    \label{eqn:normalization}
    \end{equation}
    where, $r'_{i, j}\in RSSI'_j$ is the RSSI value of $i$th AP measured at the $j$th AP. Eq. \ref{eqn:normalization} replaces no signal RSSI values with 0 and all other RSSI values in the range [0.25, 1]. Here, min = -100 and max = 0 \cite{long-term}.
    
\subsubsection{Noise Addition}\label{noise_addition}  
Our scheme randomly selects 10\% of the samples per RP and creates a Gaussian distribution curve with a standard deviation of 0.5. Random values taken from this distribution are added to the actual RSSI values as noise.

\subsubsection{Data Fitting/Model Training} Data is fitted to the \textit{LGBM} classifier model  \cite{base-lgbm}  built with tuned hyperparameters. 

\subsection{Online Phase}
The proposed scheme receives online queries of form $ RSSI''_p = [r''_{1, p}, r''_{2, p}, r''_{3, p}, \ldots, r''_{n-1, p}, r''_{n, p}]$ from a user who wants to know their location, where $r''_{i, p}$ is the RSSI value of the $i$th AP from the user's position $p$, and $p$ is unknown. The localization model predicts the location $l''_p$ (the position of the closest RP to the user's position $p$) for incoming queries and saves them to create an online database, a collection of labeled records $[RSSI''_p, l''_p]$ corresponding to the online queries.

\subsection{Malicious AP Detection}
The proposed scheme separates online and offline data samples based on reference points and conducts two tests on the RSSI values of every AP. Let, 
    
\noindent 
$A_j$ = data samples $RSSI''_j$ for RP $j$ in the online database

\noindent $A_{i,j}$ = collection of $r''_{i,j}$ for an AP $i$ where $r''_{i,j} \in RSSI''_j$ 

\noindent 
$B_j$ = data samples $RSSI'_j$ for RP $j$ in the offline database

\noindent $B_{i,j}$ = collection of $r'_{i,j}$ for an AP $i$ where $r'_{i,j} \in RSSI'_j$

\subsubsection{Check Variance} It checks the variance of an AP for every RP in the online database. If the maximum variance value of an AP equals or exceeds the variance threshold, the AP is marked as malicious. The variance threshold $TH_1$ is set to 0.05, as nearly 99\% of the variance for each AP in the offline database is within 0.05 for the UJI dataset \cite{long-term}. Let,

    \noindent $t$ = $|A_{i,j}|$ = number of entries in $A_{i,j}$
    
    \noindent $\mu$ = $\frac{1}{t} \sum r''_{i,j}$ = mean of RSSI values of an AP $i$ where $r''_{i,j} \in A_{i,j}$ 
    
    \noindent $\sigma^2_{i,j}$  = $\frac{1}{t}\sum (r''_{i,j} - \mu)^2$ = variance of an AP $i$ for RP $j$

    \noindent An AP $i$ is malicious if $\max(\sigma^2_{i, j}) \geq TH_1$, where, $j\in 1\ldots r$

    \subsubsection{Check Mean Difference} It checks the difference in the mean value of offline and online data for a specific AP for every RP. If the minimum value of the differences exceeds a defined threshold value $TH_2$, the AP is marked as malicious. We empirically set $TH_2$ to 0.005 as it detects nearly 98\% malicious APs for the UJI dataset \cite{long-term}. Let,

    \noindent $u$ = $|B_{i,j}|$ = number of entries in $B_{i,j}$ 
    
    \noindent $v$ = $|A_{i,j}|$ = number of entries in $A_{i,j}$ 

    \noindent $\frac{1}{u} \sum r'_{i,j}$ = mean of RSSI values of an AP $i$ where $r'_{i,j} \in B_{i,j}$

     \noindent $\frac{1}{v} \sum r''_{i,j}$ = mean of RSSI values of an AP $i$ where $r''_{i,j} \in A_{i,j}$

    \noindent $\delta_{i,j}$ = $(\frac{1}{u} \sum r'_{i,j} - \frac{1}{v} \sum r''_{i,j})$ = difference of mean values.
    
   \noindent An AP $i$ is malicious if  $min(\delta_{i,j}) > TH_2$, where $j\in 1\ldots r$

We considered 1560 data samples from the first month of the UJI dataset \cite{long-term} to determine $TH_1$ and $TH_2$.

\subsection{Malicious Effect Mitigation}
\subsubsection{Train Regression Model}\label{train_reg_model} 
Suppose the malicious AP detection module identifies $p$ malicious APs, leading to $m$ = $n$ - $p$ APs behaving honestly. An online query for $i$th RP can be rewritten as $RSSI''_{i} = [r''_{1, i}, r''_{2, i}, \ldots, r''_{m-1, i}, r''_{m, i}, r''_{ m+1, i}, \ldots, r''_{m+p-1, i}$, $r''_{m+p, i}]$ 
where $r''_{m+1, i} \sim r''_{m+p, i}$ are RSSI entries for malicious APs. The proposed scheme deploys $p$ LGBMRegressors \cite{base-lgbm} and trains them with offline database's reliable data 
$RSSI'_{i}$ of form $[r'_{1, i}, r'_{2, i}, \ldots, r'_{m-1, i}, r'_{m, i}]$ (contains the RSSI entries for $m$ honest APs), where $i \in 1, 2, \ldots r$.

\subsubsection{Update Online Queries and Offline Database}
The LGBMRegressors predicts $p$ RSSI values when input data is $RSSI'_{i}$ of form $[r'_{1, i}, r'_{2, i}, \ldots, r'_{m-1, i}, r'_{m, i}]$ and use these data to replace the values of malicious APs $r'_{ m+1, i}, \ldots, r'_{m+p-1, i}$, $r'_{m+p, i}$ in the offline $RSSI'_{i}$ record. Similarly, when the input data is $RSSI''_{i}$ of form $[r''_{1, i}, r''_{2, i}, \ldots, r''_{m-1, i}, r''_{m, i}]$, the predicted $p$ RSSI values replace $r''_{ m+1, i}, \ldots, r''_{m+p-1, i}$, $r''_{m+p, i}$ in the online query $RSSI''_i$. The localization model is retrained with the updated offline database. The updated query $RSSI''_i$ is used by the classification model to predict the location. The process of predicting $p$ RSSI values is inspired by Montoliu et al. \cite{impute}. The malicious AP detection module is executed after a fixed interval determined by the online database size, number of users, time to serve a location query and the number of requests/minute. If a malicious AP is detected, then the malicious effect mitigation module is executed.

\section{Experimental Evaluation}\label{sec:exp_result}
We implemented the proposed scheme using Python and evaluated its performance using the UJI's dataset \cite{long-term}. The UJI's dataset \cite{long-term} is the longest public dataset with 103584 WiFi fingerprints. We used only the data of floor 3 for brevity, as also done in \cite{31}. This data set contains 4320 data samples for training purposes for floor 3 \cite{uji-base-paper}. We incorporated noise-added samples in this set and extended its size to 4800. Initially, there were 620 APs in the environment. However, we took only 40 APs after correlated AP selection. We trained our localization model which is an LGBMClassifier \cite{base-lgbm}, with the offline dataset. We used seven performance metrics shown in Table \ref{tab:eval_metrics} to evaluate the performance. The parameters used in defining performance metrics are shown in Table \ref{tab:descriptions}. 
\begin{table}
\centering
\caption{Parameter used in Performance Metrics}
\label{tab:descriptions}
\scalebox{0.85}{
\begin{tabular}{|c|c|}
\hline
\textbf{Parameter} & \textbf{Description} \\
\hline
True Positive (TP) & Correctly predicted positive/malicious instances \\
\hline
True Negative (TN) & Correctly predicted negative/nonmalicious instances \\
\hline
False Positive (FP) & Incorrectly predicted positive/malicious instances \\
\hline
False Negative (FN) & Incorrectly predicted negative/nonmalicious instances \\
\hline
\end{tabular}}
\end{table}

\begin{table}
\centering
\caption{Evaluation Metrics}
\label{tab:eval_metrics}
\scalebox{0.85}{
\begin{tabular}{|p{2cm}|p{3.5cm}|p{3cm}|}
\hline
False Positive Rate (FPR) & Ratio of legitimate APs falsely identified as malicious &     $FPR = \frac{FP}{FP+TN}$   \\ 
\hline
False Negative Rate (FNR) & Ratio of malicious APs falsely identified as legitimate &     $FNR = \frac{FN}{FN+TP}$   \\
\hline
Precision & Ratio of correctly detected malicious APs &     $Precision = \frac{TP}{TP+FP}$   \\
\hline
Recall & Ratio of correctly identified malicious APs to the total actual malicious APs &     $Recall = \frac{TP}{TP+FN}$   \\
\hline
Accuracy & Ratio of correctly identified malicious APs and legal APs & $Accuracy = \frac{TP + TN}{TP + TN + FP + FN}$ \\
\hline
F1 Score & Relationship between precision and recall & $F1 = 2 \times \frac{precision \times recall}{precision + recall}$
\\
\hline
Mean Localization Error (MLE) & Deviation of the predicted location from the actual location & 
$
MLE = \sqrt{(x - \Bar{x})^2 + (y - \Bar{y})^2}$ \\
\hline
\end{tabular}}
\end{table}

\subsection{Impact of Noise Addition}

\begin{figure}
    \centering
    \includegraphics[
    width = 0.45\textwidth
    ]{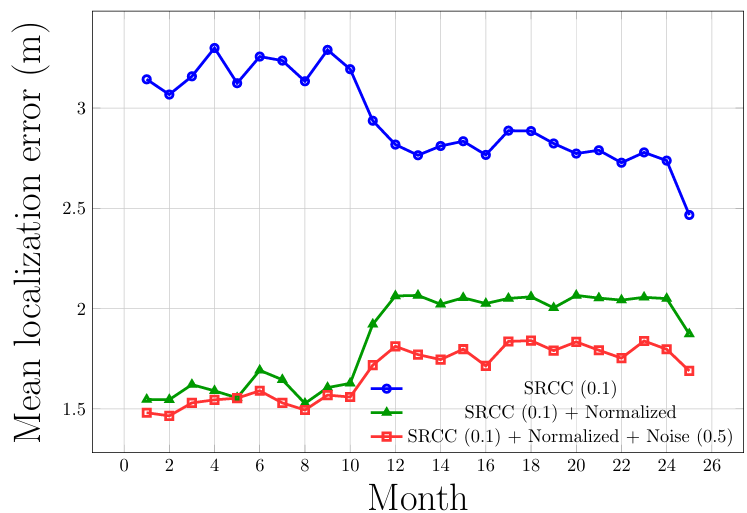}
    \caption{Impact of noise addition.}
    \label{fig:noise}
\end{figure}

Figure \ref{fig:noise} shows mean localization errors over the data of 25 months from UJI's dataset \cite{long-term}. The curve labeled \textit{SRCC (0.1)} shows the effect of correlated AP selection only, the curve labeled \textit{SRCC (0.1) + Normalized} shows the impact of AP selection and normalization, and the curve labeled \textit{SRCC (0.1) + Normalized + Noise (0.5)} shows the effect of AP selection, normalization, and noise addition. Figure \ref{fig:noise} shows a notable reduction in localization error for the curve \textit{SRCC (0.1) + Normalized + Noise (0.5)} with respect to \textit{SRCC (0.1) + Normalized}  after the 10th month because the model becomes robust to RSSI fluctuations in the long term. The MLE is reduced by nearly 16\% in the 16th month, which is the maximum change of the red curve with respect to the green one. Normalizing data helps models converge faster, and adding noise makes the model more robust against long-term noise. Correlated AP selection, noise addition, and data normalization bring the MLE of the proposed scheme below 2 meters, outperforming existing machine learning schemes \cite{31}. 

\begin{figure*}
    \centering
    \subfigure[False positive rate]{\includegraphics[width=0.28\textwidth]{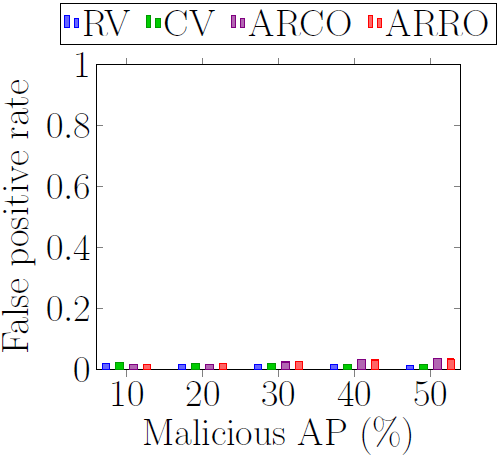}
    \label{fig:indiv_fpr}}
    \hfil
    \subfigure[False negative rate]{\includegraphics[width=0.28\textwidth]{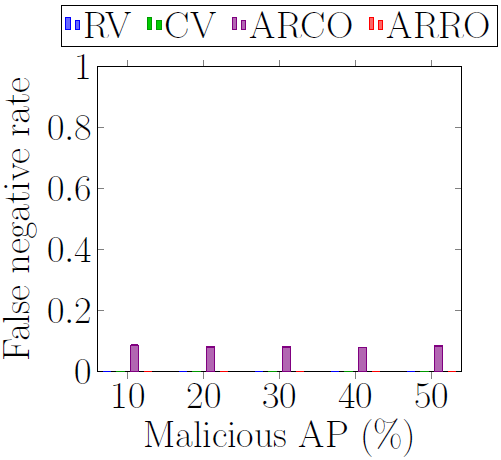}
    \label{fig:indiv_fnr}}
    \subfigure[Recall]{\includegraphics[width=0.28\textwidth]{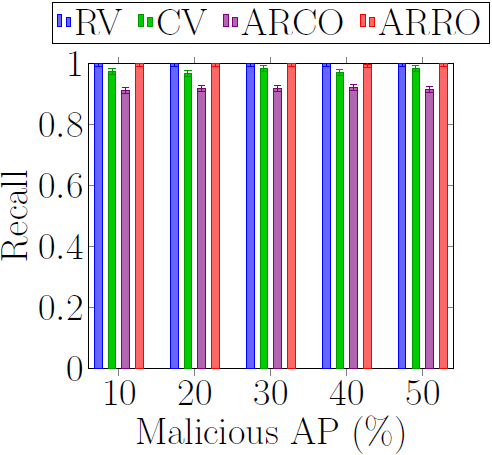}
    \label{fig:indiv_recall}}
    \subfigure[Precision]{\includegraphics[width=0.28\textwidth]{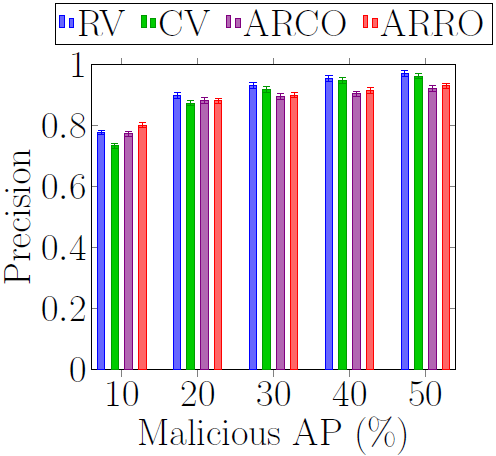}
    \label{fig:indiv_precision}}
    \subfigure[Accuracy]{\includegraphics[width=0.28\textwidth]{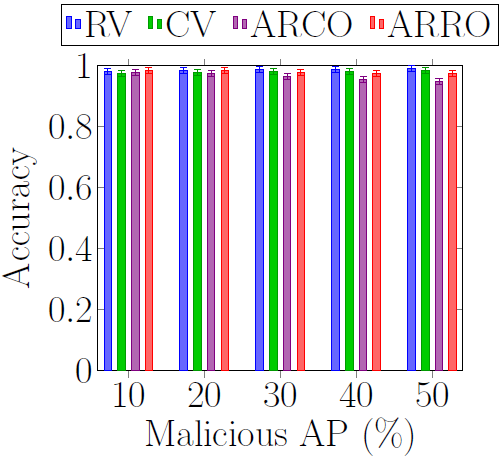}
    \label{fig:indiv_accuracy}} 
    \subfigure[F1 score]{\includegraphics[width=0.28\textwidth]{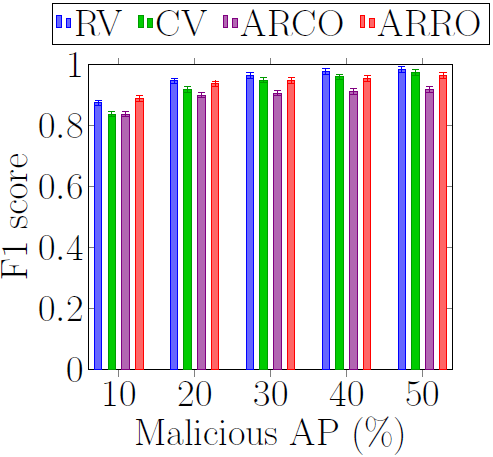}
    \label{fig:indiv_f1score}}
    \caption{Performance analysis of the proposed scheme.}
    \label{fig: result_charts}
\end{figure*}

\subsection{Malicious AP Detection}
We considered four malicious behaviors of an AP following the studies of \cite{37}, \cite{35}, \cite{8}, \cite{veremi}: \textit{constant RSSI value}, \textit{random RSSI value}, \textit{constant offset alteration of the actual RSSI value}, and \textit{random offset alteration of the actual RSSI value} as shown in Table \ref{tab:attacts}. We evaluated the malicious AP detection capability of the proposed scheme using test data comprising 1560 samples from the first month's data (floor 3). We chose this data because the first month's data is less susceptible to errors caused by undetected malicious APs and environmental changes. For each attack type, each experiment was iterated 100 times with malicious APs in the range $10\%\sim50\%$. Each data point in the result was averaged over 100 iterations. 
\begin{table}
\centering
\caption{Attack Description and Acronyms}
\label{tab:attacts}
\scalebox{.85}{
\begin{tabular}{|p{2cm}|c|p{4.5cm}|}
\hline
\textbf{Attack Type} & \textbf{Acronym} & \textbf{Description} \\
\hline
Constant Value & \textbf{CV} & A constant RSSI value is set for the malicious AP. Here, the constant value is 100, interpreted as no signal from the AP. It also represents that an AP is removed.\\
\hline
Random Value & \textbf{RV} & Random RSSI values are assigned for APs. \\
\hline
Actual RSSI Plus Constant Offset & \textbf{ARCO} & A fixed offset ([-100,0]) is added to
the actual RSSI value for malicious APs. This also represents the movement of AP from its original position. \\
\hline
Actual RSSI Plus Random Offset & \textbf{ARRO} & A random offset ([-100,0]) is taken and added to the actual RSSI value. \\
\hline
\end{tabular}}
\end{table}

\subsubsection{Performance Analysis of the Proposed Scheme}\label{performance_proposed_scheme}
\begin{figure*}
    \centering
    \subfigure[False positive rate]{\includegraphics[width=0.28\textwidth]{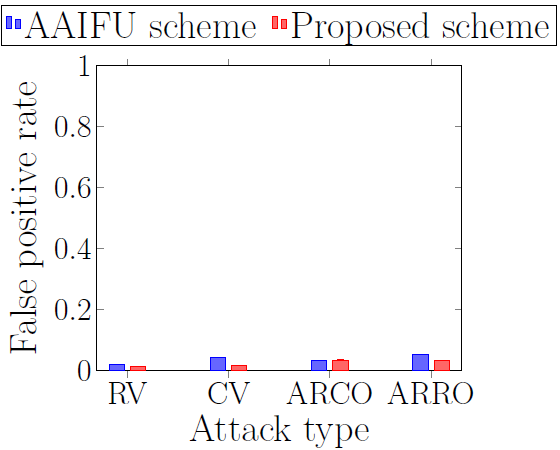}
    \label{fig:comp_fpr}}
    \subfigure[False negative rate]{\includegraphics[width=0.28\textwidth]{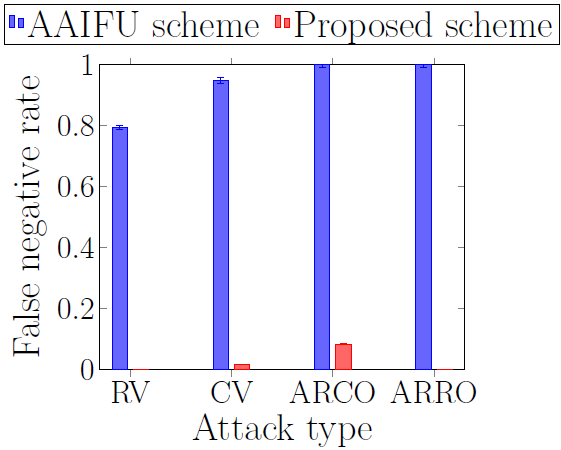}
     \label{fig:comp_fnr}}
    \subfigure[Recall]{\includegraphics[width=0.28\textwidth]{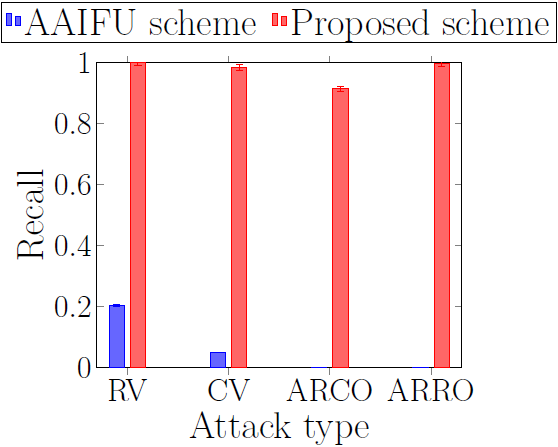}
     \label{fig:comp_recall}}
    \subfigure[Precision]{\includegraphics[width=0.28\textwidth]{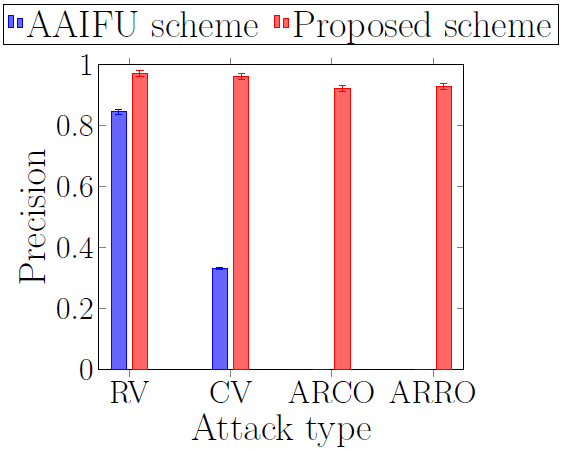}
    \label{fig:comp_precision}}
    \subfigure[Accuracy]{\includegraphics[width=0.28\textwidth]{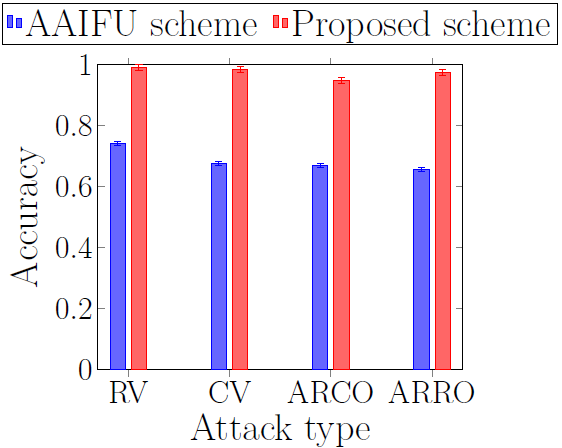}
     \label{fig:comp_accuracy}} 
    \subfigure[F1 score]{\includegraphics[width=0.28\textwidth]{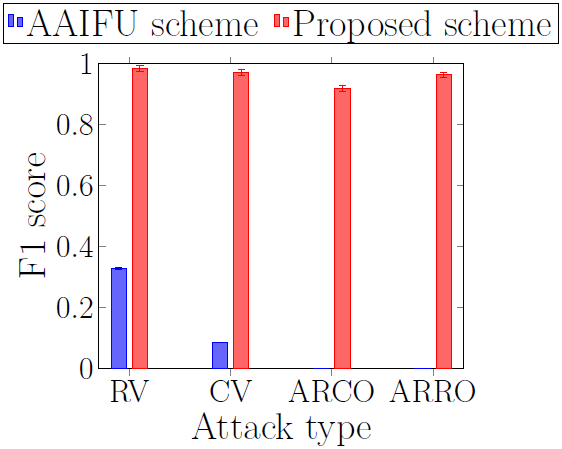}
     \label{fig:comp_f1score}}
    \caption{Performance comparison between AAIFU and proposed schemes.}
    \label{fig: comp_result_charts}
\end{figure*}
Figure \ref{fig: result_charts} shows the performance of the proposed detection module.
\begin{enumerate}

\item\textit{False Positive Rate (FPR):} Figure \ref{fig:indiv_fpr} shows that the FPR consistently remains below 4\%. The legitimate APs sometimes exhibit signal variation due to temporal effects and are marked as malicious by the detection system.  
     
    \item\textit{False Negative Rate (FNR):}\label{test_fnr} Figure \ref{fig:indiv_fnr} shows that, except for the ARCO attack, all other attacks show a negligible FNR. As marking an AP as malicious in our experiment was a random process, in very few cases, the selected malicious AP was assigned the same original RSSI value, leading to a false negative outcome. In ARCO attacks, FNR stays nearly 10\%. In this attack, we add or subtract a fixed value and also add an offset in the noise addition module, leading to very little change in the mean that can not be distinguished with the mean difference checker, resulting in an increased FNR. 
    
    \item\textit{Recall:} The recall is consistently 100\% for RV and ARRO attacks, almost 98\% for CV attacks, and nearly 92\% for ARCO attacks, as shown in Fig. \ref{fig:indiv_recall}. In ARCO attacks, the recall is lower for the higher FNR than for other attacks. 
    
    \item\textit{Precision:} Figure \ref{fig:indiv_precision} shows that the detection rate increases with escalating malicious APs for all attack types. The proposed scheme achieves a satisfactory precision of nearly 98\%, 97\%, 95\%, and 96\% for the RV, CV, ARCO, and ARRO attacks, respectively, for 50\% malicious APs. 
    
    \item\textit{Accuracy:} Figure \ref{fig:indiv_accuracy} shows that the accuracy of the detection module consistently remains above 95\% for all types of attack, underscoring the robustness and reliability of the detection system across various scenarios.
    
    \item\textit{F1 Score:} Figure \ref{fig:indiv_f1score} shows that the F1 score increases with malicious APs for all attack types. The proposed scheme exhibits an F1 score within 84\%$\sim$99\%. 
\end{enumerate}

\subsubsection{Comparative Analysis of AAIFU and Proposed Schemes}
In this section, we compare the performance of the AAIFU \cite{35} and proposed schemes for 50\% malicious APs. 
\begin{enumerate}
\item \textit{False positive rate (FPR):} Figure \ref{fig:comp_fpr} shows that the FPR ranges between 2\%$\sim$6\% and 2\%$\sim$4\% for AAIFU and proposed schemes, respectively. Due to the temporal effect, both schemes exhibit FPR.

\item\textit{False negative rate (FNR):} Figure \ref{fig:comp_fnr} shows that the FNR in the AAIFU scheme is dramatically higher, ranging between 80\%$\sim$100\%. The proposed scheme never misses any malicious AP in RV and ARRO attacks. It shows nearly 9\% and 2\% for ARCO and CV attacks, respectively. In the AAIFU scheme, FNR is visible when the alarm frequency of altered APs is lower than that of unaltered APs. From the alarm frequency distribution, the AAIFU scheme uses a clustering method to find the actual altered APs, ignoring the altered (malicious) APs with low alert frequency, causing a significant FNR. 

\item\textit{Recall: }The AAIFU scheme exhibits a recall of nearly 0\% for ARRO and ARCO attacks, visible from Fig. \ref{fig:comp_recall}. For RV and CV attacks, the AAIFU scheme shows lower recall values of around 21\% and 5\%, respectively. On the other hand, the proposed scheme exhibits a recall of 100\% for RV and ARRO attacks and always maintains a recall $\geq$ 92\% for other attacks. The ineffective clustering of the AAIFU scheme lead to higher FNR and 0 TPs. 

\item{Precision:} As shown in Fig \ref{fig:comp_precision}, the AAIFU scheme shows a precision of nearly 0\% for ARRO and ARCO attacks due to 0\% TP for nearly 100\% FNR. It displays a precision of approximately 85\% and 34\% for RV and CV attacks, respectively. In contrast, the proposed scheme shows precision values $\geq$ 93\% for all attack types.  

\item{Accuracy:} Figure  \ref{fig:comp_accuracy} shows that the highest accuracy for the AAIFU and proposed schemes are approximately 75\% and 99\%, respectively. Despite the lower TP, the AAIFU scheme exhibits a higher TN value, resulting in greater accuracy than recall and precision. 

\item{F1 score:} The AAIFU scheme exhibits an F1 score of 0\% for ARCO and ARRO attacks and approximately 33\% and 9\% F1 scores for RV and CV attacks, respectively, visible from Fig \ref{fig:comp_f1score}. On the contrary, the proposed scheme maintains a good balance between precision and recall, with an F1 score $\geq$ 92\% for all attack types. The 0 in precision and recall results in 0 in the F1 score for the AAIFU scheme. 
\end{enumerate}

\subsection{Impact of Malicious  Effect Mitigation}
Five Cumulative Distribution Function (CDF) curves in Fig. \ref{fig:result_CDF} show the localization error using 8112 samples for 25 months from the UJI dataset \cite{long-term} for the following scenarios:
\begin{figure*}
    \centering
    \subfigure[RV]{\includegraphics[width=0.24\textwidth]{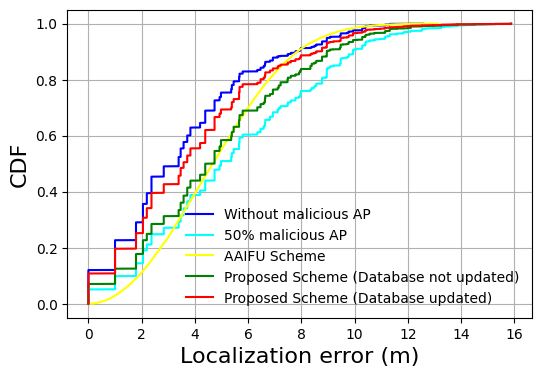}} 
    \subfigure[CV]{\includegraphics[width=0.24\textwidth]{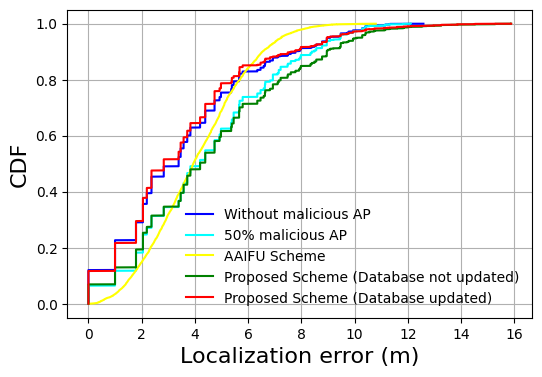}}
    \subfigure[ARCO]{\includegraphics[width=0.24\textwidth]{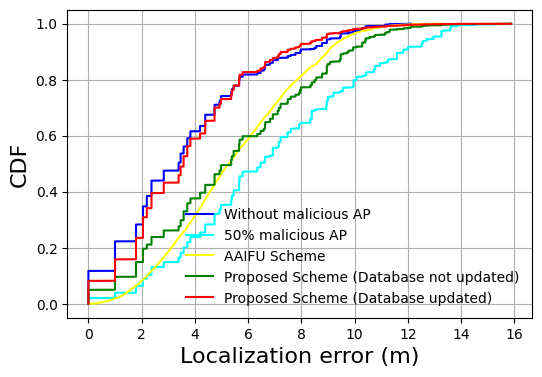}}
    \subfigure[ARRO]{\includegraphics[width=0.24\textwidth]{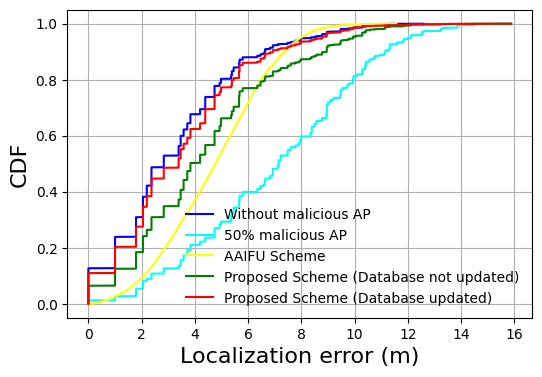}}
    \caption{CDF curves of localization error for different attacks.}
    \label{fig:result_CDF}
\end{figure*}
\begin{enumerate}[left=0pt]
    \item \textit{Without malicious AP} -- shows the MLE without the presence of malicious AP. Here, the malicious AP detection and effect mitigation modules were not deployed.

    \item \textit{50\% malicious AP} -- presents the MLE when 50\% of the APs are malicious. Here, malicious AP detection and effect mitigation modules were not used.

   \item \textit{AAIFU Scheme} -- shows the MLE for the AAIFU scheme \cite{35}, where 50\% APs were malicious.
    \item \textit{Proposed Scheme without offline database updation and model retraining} -- displays the MLE for the proposed scheme where a malicious AP detection module was used and online queries were updated. Here, the offline database was not updated, the model was not retrained, and 50\% of the APs were malicious.
    
    \item \textit{Proposed scheme with offline database updation and model retraining} -- shows the MLE for the proposed scheme where the malicious AP detection and effect mitigation modules were used. The offline database was updated, the model was retrained, and 50\% of the APs were malicious.
\end{enumerate}
Figure. \ref{fig:result_CDF} shows that the curve for 50\% malicious AP exhibits significant degradation from the curve with no malicious AP across all types of attacks. The proposed scheme, which includes database update and model retraining, shows an almost unnoticeable deviation from the curve with no malicious AP for every type of attack. However, the proposed scheme that does not include the database updation and model retraining exhibits degraded performance compared to the proposed scheme with the database updation curve. By updating the database and retraining the model after identifying malicious APs, the localization error is minimized. The AAIFU scheme \cite{35} focuses on the steady change in the behavior of APs rather than the dynamic malicious effects during the online phase. This scheme produces results similar to our proposed scheme without requiring the database update, as shown in Fig. \ref{fig:result_CDF}. The CDF curve for the AAIFU scheme is smoother for the use of a regressor, and the CDF curve for the proposed scheme appears more like a staircase for the use of a classifier.

 It is observed from experiments that the proposed scheme effectively handles at most 60\% malicious APs. After that, the performance deviates to a greater extent. 

\subsection{Execution Time}
\label{experimentsummary}

\begin{table*}
    \centering
    \normalsize
    \caption{Execution time (seconds) of the AAIFU and proposed scheme for 10\%$\sim$50\% malicious AP}
    \scalebox{.60}{
    \begin{tabular}{lccccc|ccccc}
        \toprule
        \multirow{2}{*}{\textbf{Step}} & \multicolumn{5}{c|}{AAIFU scheme} & \multicolumn{5}{c}{Proposed scheme} \\
        \cmidrule(lr){2-6} \cmidrule(lr){7-11}
        & 10\% & 20\% & 30\% & 40\% & 50\% & 10\% & 20\% & 30\% & 40\% & 50\% \\
        \midrule
        Detection & 599.69 & 604.14 & 575.38 & 602.94 & 554.47 & 1.43 & 1.39 & 2.19 & 1.77 & 1.32 \\
        Database updation & 3.46 & 1.76 & 1.92 & 0.72 & 0.95 & 8.38 & 6.85 & 13.84 & 13.13 & 17.30 \\
        Model retraining & 0.91 & 0.67 & 1.05 & 0.77 & 0.65 & 8.56 & 12.19 & 9.50 & 10.34 & 10.67 \\
        Inference & 0.05 & 0.05 & 0.07 & 0.08 & 0.04 & 9.05 & 9.13 & 9.07 & 9.38 & 9.49 \\
        \bottomrule
        Total & 604.11 & 606.61 & 578.42 & 604.51 & 556.12 & 27.42 & 29.56 & 34.61 & 34.62 & 38.78 \\
        \bottomrule
    \end{tabular}}
    \label{tab:run_time}
\end{table*}

The proposed scheme exhibits a gradual increase in runtime with the increasing number of malicious APs, as shown in Table \ref{tab:run_time}. The AAIFU scheme randomly detects malicious APs, which is displayed by the non-gradual increase of runtime. The AAIFU scheme spends a significant amount of time in the detection phase than the proposed scheme, as the proposed scheme uses only some simple statistical calculations, whereas the AAIFU scheme uses GBDTRegressors. The AAIFU scheme detects fewer APs than the proposed scheme. Hence, even though both schemes use regressors for updating database, the proposed scheme takes more time than the AAIFU scheme. The proposed scheme takes more time than the AAIFU scheme in model retraining, indicating that training a classifier is more time-consuming than a regressor. Lastly, in the inference phase, the proposed scheme involves regressors to predict the values of malicious APs in online queries, resulting in greater time consumption compared to the AAIFU scheme. The overall time consumed by the proposed scheme is approximately 94\% less than the AAIFU scheme. 
\section{Conclusion}\label{conclusion}
We have proposed a reliable and efficient long-term indoor localization system by incorporating malicious AP detection and effect mitigation modules. Experimental results show that the proposed scheme achieves detection accuracy above 95\% for all attack types. The malicious effect mitigation module ensures the accurate functioning of the indoor localization system by restoring the system's performance to the initial state, where the system operates without any malicious APs. Besides, the proposed indoor localization system cuts down the execution time by approximately 94\% compared to existing works. We believe the proposed scheme enhances the existing literature by offering a faster and more reliable indoor localization system. We plan to extend this work by considering device diversity, attack scenarios for online and offline databases, system scalability, and generalizability.

\bibliography{main} 
\bibliographystyle{ieeetr}

\end{document}